\documentclass[aip,jcp,unsortedaddress,preprint]{revtex4-1}

\newcommand{\figwidth}{0.79 \linewidth}

\usepackage{graphicx}
\usepackage[normalem]{ulem}
\usepackage{color}
\usepackage{amsmath,amssymb}
\usepackage[T1]{fontenc}

\newcommand{\nddp}{ND$_3^+$}
\newcommand{\nhdp}{NH$_3^+$}

\newcommand{\nhtp}{NH$_2^+$}
\newcommand{\ndd}{ND$_3$}
\newcommand{\nhd}{NH$_3$}

\newcommand{\nddps}{ND$_3^+$ }
\newcommand{\nhdps}{NH$_3^+$ }
\newcommand{\ndtps}{ND$_2^+$ }
\newcommand{\nhtps}{NH$_2^+$ }
\newcommand{\ndds}{ND$_3$ }
\newcommand{\nhds}{NH$_3$ }

\newcommand{\jk}[2]{$\mathrm{#1}_{\mathrm{#2}}$}
\newcommand{\AAs}{\AA\hspace{1mm} }

\begin{document}
\title{Dynamics of gas phase $\mathrm{Ne}^*$ + \nhds and $\mathrm{Ne}^*$ + \ndds Penning ionisation at low temperatures }

\author{Justin Jankunas}
\author{Benjamin Bertsche}
\affiliation{Institute for Chemical Sciences and Engineering, Ecole Polytechnique F\'ed\'erale de Lausanne, 1015 Lausanne, Switzerland}
\author{Krzysztof Jachymski}
\affiliation{Faculty of Physics, University of Warsaw, Ho{\.z}a 69, 00-681 Warsaw, Poland}
\author{Micha\l\ Hapka}
\affiliation{Faculty of Chemistry, University of Warsaw, Pasteura 1, 02-093 Warsaw, Poland}
\author{Andreas Osterwalder}
\email[]{andreas.osterwalder@epfl.ch}
\affiliation{Institute for Chemical Sciences and Engineering, Ecole Polytechnique F\'ed\'erale de Lausanne, 1015 Lausanne, Switzerland}

\date{\today}

\begin{abstract}
Two isotopic chemical reactions, $\mathrm{Ne}^*$ + \nhd, and $\mathrm{Ne}^*$ + \ndd, have been studied at low collision energies by means of a merged beams technique. 
Partial cross sections have been recorded for the two reactive channels, namely $\mathrm{Ne}^*$ + \nhd $\rightarrow$ Ne + \nhdp$+e^-$, and $\mathrm{Ne}^*$ + \nhd $\rightarrow$ Ne + \nhtp + H$+e^-$, by detecting the \nhdps and \nhtps product ions, respectively. 
The cross sections for both reactions were found to increase with decreasing collision energy, $E_{coll}$, in the range 8 $\mu$eV$<E_{coll}<$ 20 meV.
The measured rate constant exhibits a curvature in a log(k)-log($E_{coll}$) plot from which it is concluded that the Langevin capture model does not properly describe the $\mathrm{Ne}^*$ + \nhds reaction in the entire range of collision energies covered here. 
Calculations based on multichannel quantum defect theory were performed to reproduce and interpret the experimental results.
Good agreement was obtained by including long range van der Waals interactions combined with a 6-12 Lennard-Jones potential.
The branching ratio between the two reactive channels, $\Gamma = \frac{[NH_2^+]}{[NH_2^+]+[NH_3^+]}$, is relatively constant, $\Gamma\approx 0.3$, in the entire collision energy range studied here. 
Possible reasons for this observation are discussed and rationalised in terms of relative time scales of the reactant approach and the molecular rotation.
Isotopic differences between the $\mathrm{Ne}^*$ + \nhds and $\mathrm{Ne}^*$ + \ndds reactions are small, as suggested by nearly equal branching ratios and cross sections for the two reactions. 
\end{abstract}
\maketitle

\section{INTRODUCTION}
Cold and ultracold molecular collisions remain a largely unexplored area of gas phase reaction dynamics. 
Traditional gas phase reaction dynamics experiments typically access collision energies above 50 K, and thus probe mainly the repulsive part of the intermolecular potentials. 
Low energy collisions complement our understanding of these dynamics since they are sensitive in particular to the long-range attractive part of these potentials.  
The combined knowledge of chemical reactions at high and low temperature will thus provide a more complete picture of bimolecular collisions. 
Recent advances in the translational control of neutral molecules\cite{Krems:2009wc,vandeMeerakker:2012ft,PSJ2012} finally permit to venture into the regime of cold molecular collisions.
The study of scattering at collision energies ($E_{coll}$) in the range below 10 K is interesting for several reasons. 
As $E_{coll}$ approaches zero, the number of partial waves that contribute to the reaction diminishes, highlighting the role of possible shape resonances, 
which so far have been observed experimentally only in a handful of cold chemical reactions at collision energies below a few Kelvin. \cite{Henson:2012kr,Chefdeville:2013jc,2012PhRvL.109b3201C}
In a related study the hydrogen abstraction reaction OH + CH$_3$OH$\rightarrow$H$_2$O+CH$_3$O was found to have a surprisingly large reaction rate below 60 K,\cite{Shannon:2013hn} and this was attributed to quantum mechanical tunnelling through the activation barrier. 
As first suggested by Wigner,\cite{Wigner:1948bw} the rate constants of certain chemical reactions tend to a constant value when only $\ell$ = 0 partial waves (s-waves), that have no centrifugal barrier, contribute to the reaction. 
Calculations carried out for F + H$_2$, H + HCl, and Li + HF reactions, for example (see Tables 3.1 and 3.3 of Ref.  \onlinecite{Niehaus:1981vb}), confirm a non-zero rate constant at T = 0 K. 
Verification of these predictions remains an experimental challenge. 

Until recently the only experiment that yielded collision energies below 50 K was an experiment called \emph{Cin\'etique de R\'eaction en Ecoulement Supersonic Uniforme} (CRESU).\cite{Sims:1995wz} 
Here, one reactant and a precursor to the second one are co-expanded in a uniform expansion through a Laval nozzle. 
The reaction is induced by the formation of the second reactant by electron bombardment or with a laser. 
Recent advances in the production of cold and ultracold neutral molecules have opened up ways to reach the low temperature regime\cite{Krems:2009wc} but to date only few reactive collisions have been studied. 
Very recently, crossed beam experiments have been performed at low crossing angle, pushing the collision energy to less than 5 K.\cite{Chefdeville:2013jc,2012PhRvL.109b3201C}
In order to study low-temperature ion-molecule reactions Willitsch and co-workers have combined methods to prepare translationally cold molecules with so-called Coulomb crystals where cations are stored at $\mu$K translational temperatures.\cite{Willitsch:2012uq}
A recent combination of this experiment with a device to select molecular conformers based on their electric dipole moments revealed markedly different reactivities for rotamers of 3-aminophenol (C$_6$H$_7$NO) in collisions with cold Ca$^+$ atoms in a Coulomb crystal.\cite{Rosch:2014uq,Chang:2013dm}
Parazzoli et al. overlapped a sample of trapped ammonia molecules with magneto-optically trapped Rb atoms to investigate the effects of electric and magnetic fields on neutral reactions.\cite{Parazzoli:2011ts}
The effects of electric fields on collisions between two polar molecules were studied experimentally by crossing magnetically trapped OH radicals with a beam of velocity-filtered, cold ammonia molecules.\cite{Sawyer:2011bh}
Good agreement with theoretical calculations was found in both of these studies.
Ultracold collisions between KRb molecules and K-atoms or Rb-atoms were studied experimentally, and theoretically by multichannel quantum defect theory (MQDT).\cite{Ospelkaus:2010cb}
By nature of the experiment these studies provided complete control over the internal state of the molecule which was prepared in a combination of magnetic fields and Raman-pumping to reach different selected states.
DeMiranda et al. took the same experiment an important step further and studied the rate of the bimolecular reaction KRb + KRb $\rightarrow$ K$_2$ + Rb$_2$ in the $\mu$K regime.\cite{deMiranda:2011gd}
A reduction of nearly two orders of magnitude in the reaction rate was observed when the two KRb molecules collided in a side-to-side fashion. 
Collisions of $^6$Li atoms with associated $^6$Li$_2$ molecules in the ultracold regime showed marked deviations from the universal predictions based on van der Waals interaction.\cite{Idziaszek2010,Wang:2013de}

In molecular beams experiments, a conceptually different approach is required to reach collision energies below 5 K: merged molecular beams.
This technique, developed for ions several decades ago,\cite{Phaneuf:1999ga} was first demonstrated on neutrals by Narevicius et al.\cite{Henson:2012kr}, and more recently also at EPFL.\cite{Bertsche:2014ub,jankunas:arxiv}
The key idea behind the merged beams method is that by studying collisions in two molecular beams that are merged into a single beam, the collision angle is zero and the relative velocity of the reactants in the moving frame of reference becomes simply the difference between the two beam speeds. 
In contrast to a crossed beams experiment this can become zero even at high beam velocities, thus rendering the slowing of the molecules unnecessary. 

\begin{figure}
 \includegraphics[width=3 in]{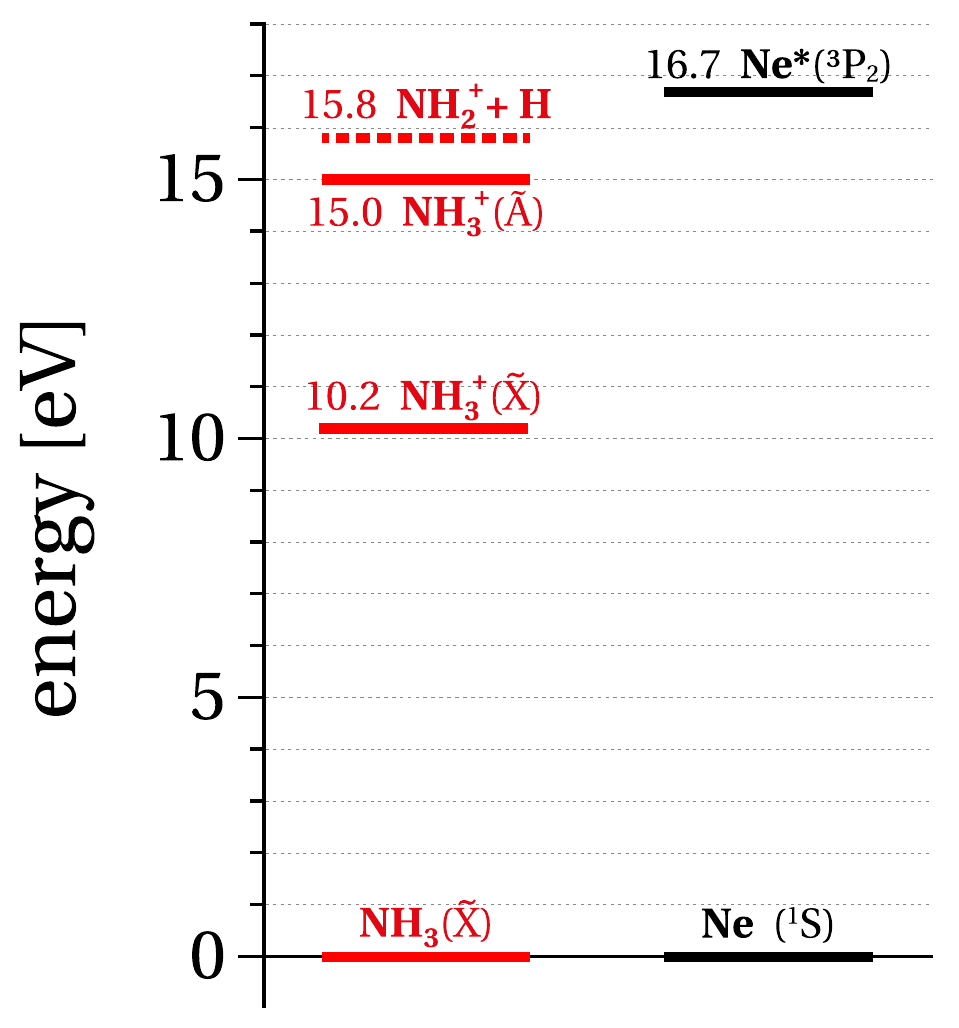}%
\caption{\label{levels}Energy level scheme showing relevant energy levels for neon (right) and ammonia (left). In the present experiment Penning ionisation of ammonia in collisions with Ne($^3P_2$) is studied, producing \nhdps and \nhtps by ionising through the $\tilde{\mathrm{X}}$-state and the $\tilde{\mathrm{A}}$-state, respectively.}
 \end{figure}
Penning ionization (PI) is an important elementary electron transfer process that is an ideal system to investigate low temperature chemistry.\cite{Henson:2012kr}
In PI the collision of a metastable species A* with a ground state species B leads to ionization of B and relaxation of A, according to the equation A* + B $\rightarrow$ A + B$^++e^-$. 
A wealth of data exists on the simplest type of PI where A* is an electronically excited rare gas atom and B an atom in the ground electronic state.\cite{Niehaus:1981vb,Siska:1993cf}
In this case experiment and theory are far advanced, and often exhibit excellent agreement. Theory is less advanced when B is a molecule and the intermolecular potential becomes a highly dimensional surface.
Until recently\cite{Henson:2012kr} all experimental studies of PI have been performed at $E_{coll}>$ 0.02 eV (250 K), and thus little is known about the dynamics at lower temperatures, where a theoretical description requires much more precise knowledge of the details of the interaction potential. 
The first merged beams experiments performed in the Narevicius group studied PI by He* of H$_2$, HD, D$_2$, and Ar.\cite{Henson:2012kr,Narevicius:2014te}
They observed very pronounced resonance structures that were quite accurately reproduced theoretically and in this way served to optimize the theoretical interaction potential between the reactants. 
In order to access the influence of a more complex molecular structure on the low-energy dynamics of PI we have undertaken the study of the PI of \nhds and \ndds molecules by metastable Ne($^3P_J$) (henceforth labeled $\mathrm{Ne}^*$) atoms at 100 mK $< E_{coll}/k_B <$ 250 K ($k_B$ is the Boltzmann constant) using our merged beams apparatus.\cite{Bertsche:2014ub,jankunas:arxiv} 
Figure \ref{levels} shows the relevant electronic energy levels for this reaction. 
The lowest $^3P_J$  multiplet of Ne lies $\approx$16.7 eV above the ground state.
Since the ionization potential of \nhds is only 10.2 eV, the internal energy of $\mathrm{Ne}^*$ is sufficient to ionize it. 
PI has been shown previously to be predominantely vibrationally adiabatic,\cite{Bevsek:1995gs,Siska:1993cf} and this has also been observed for the present reaction.\cite{BenArfa:1999el}

However, as shown in Fig. \ref{levels} multiple electronic states are accessible.
The \nhdps product can be formed either in the electronic ground state or in the excited $\tilde{\mathrm{A}}$ state. 
\nhdp($\tilde{\mathrm{X}}$) has the electronic configuration $(1a_1)^2(1e)^4(2a_1)^1$, corresponding to removal of an electron from the lone pair of \nhd, while for \nhdp($\tilde{\mathrm{A}}$) the configuration is $(1a_1)^2(1e)^3(2a_1)^2$ which means that an electron has been removed from one of the N-H bond orbitals. 
The missing electron from the N-H bond reduces the strength of that bond in the $\tilde{\mathrm{A}}$-state. 
Indeed, for $\mathrm{Ne}^*$+\nhds two reaction channels have been observed experimentally,\cite{BenArfa:1999el}
\begin{eqnarray}
\mathrm{Ne}^* + \mathrm{NH}_3 &\rightarrow& \mathrm{Ne} + \mathrm{NH}_3^++e^- \hspace{3cm}\mathrm{and}\label{eq:nodiss}\\
\mathrm{Ne}^* +  \mathrm{NH}_3 &\rightarrow& \mathrm{Ne} +  \mathrm{NH}_2^+ + \mathrm{H}+e^-. 	\label{eq:diss}					
\end{eqnarray}
The process described by Eq. \eqref{eq:nodiss} is believed to proceed through the $\tilde{\mathrm{X}}$ state while the second one goes through the $\tilde{\mathrm{A}}$ state.
The most detailed study of this reaction to date has been done by Ben-Arfa et al. \cite{BenArfa:1999el} who measured the cross section and branching ratio of Eqs. \eqref{eq:nodiss} and \eqref{eq:diss} at 300 K $<E_{coll}/k_B<$ 3000 K.  
An important result of that study is that the fraction of \nhtps products increases with increasing collision energy. 
It has been suggested that \nhdps reaction products (Eq. \ref{eq:nodiss}) result predominantly from $\mathrm{Ne}^*$ + \nhds collisions wherein the $\mathrm{Ne}^*$ atom approaches \nhds along the lone pair, whereas an approach along the N--H bond yields \nhtps products (Eq. \ref{eq:diss}).\cite{BenArfa:1999el,Levine:2009fb}
In the present work the behaviour of the branching ratio $\Gamma=\frac{[\mathrm{NH}_2]}{[\mathrm{NH}_2]+[\mathrm{NH}_3]}$ is studied as $E_{coll}$ approaches zero.
One might expect $\Gamma$ to approach one of its limiting values (0 or 1) at the lowest collision energy. 
The underlying hypothesis here is based on the notion of the minimum energy path (MEP).
The principal idea of the MEP model is that the dynamics of a molecular collision will follow the path on the PES along which the energy is always minimal. 
While this concept has proven very useful in the explanation of a large number of chemical reactions, in recent years several examples have been found where it does not strictly apply, or simply fails.
Most notable examples are the tunnelling reaction mentioned above\cite{Shannon:2013hn} or collisions where so-called roaming is observed.
In such cases the reaction does not proceed simply on a single path following the MEP.
Instead a transiently bound complex is formed in which one reactant "roams" around the other one before the reaction finally takes place.\cite{Townsend:2004fj,Bowman:2011jm,Bowman:bf}
The increasing de Broglie wavelengths at low energies invalidates the assumptions that the reactants follow any classical trajectory, so the MEP approach cannot give full understanding of the problem. 
In the present system the two accessible channels, and the temperature dependent value of $\Gamma$ should provide information about the applicability of the MEP concept to particular low-temperature reactions.

\section{EXPERIMENT}
\begin{figure}
 \includegraphics[width=\figwidth]{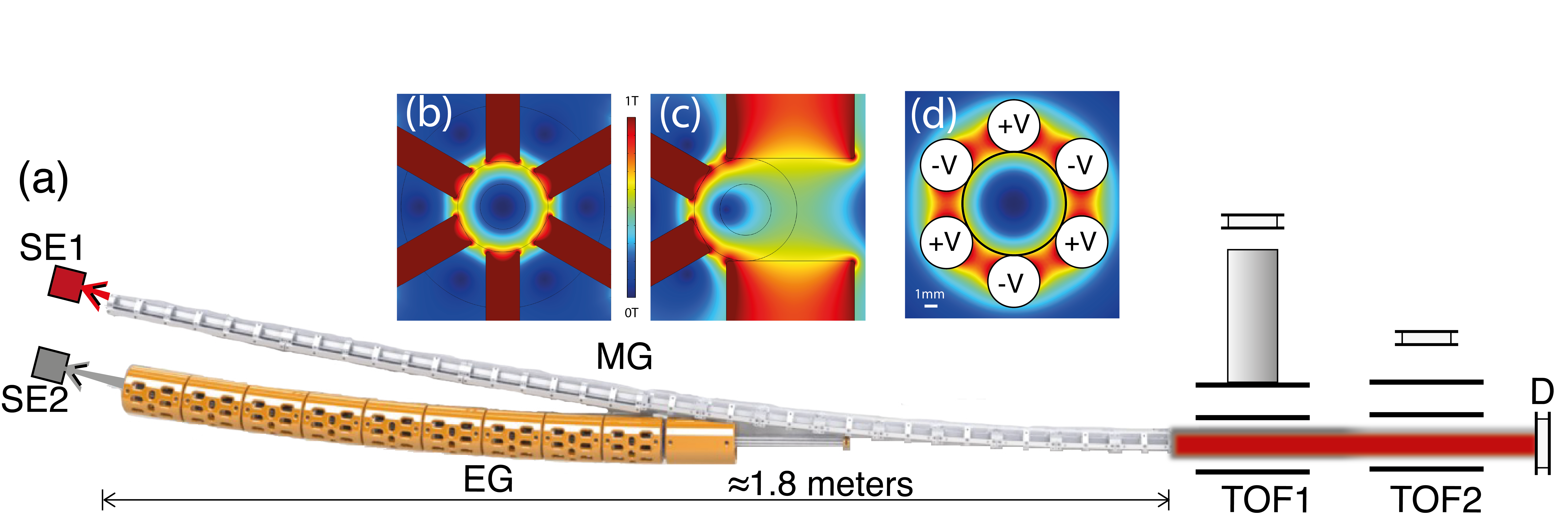}%
\caption{\label{expt}(a) Sketch of the experimental setup. SE1 and SE2 are supersonic expansions producing Ne$^*$ and \nhd, respectively. MG and EG are a magnetic and electric guide, respectively. Reaction products are detected using the time-of-flight mass spectrometer TOF1, and beam density normalisation is done using TOF2 for \nhds and D for Ne$^*$. (b) Cross-section through the first section of the MG, showing the magnetic field distribution in the hexapole guide. (c) Cross section through the MG in the second section showing the quadrupole configuration. (d) Cross section through the EG, showing the electric field distribution.}
 \end{figure}
The experimental setup is shown in Fig. \ref{expt}.
The underlying principle of the merged beam technique in reducing the collision energy is to bend one or two molecular beams, and to perfectly overlap them in space and time. In a traditional crossed beam experiment the collision energy is given as 
\begin{equation}
E_{coll}=\frac{\mu(|\vec{v}_1|^2+|\vec{v}_2|^2-2|\vec{v}_1||\vec{v}_2|\cos(\theta))}{2},
\end{equation}
where $\mu$ is the reduced mass, the $\vec{v}_i$ are the beam velocities, and $\theta$ is the angle between the two beams. 
Reaching $E_{coll}=0$ requires to either set both beam velocities to zero, or to set $\theta=0$ and $|\vec{v}_1|=|\vec{v}_2|$. 
The second option is the working principle of the merged beams experiment. 
Equal beam velocities are achieved by appropriately choosing the conditions to produce the molecular beams. 
Zero angle between the beams requires at least one beam to be bent onto the axis of the other one. 
This is achieved by using a curved electric or magnetic guide that captures the molecular beam at its source and confines it transversely while adjusting its angle relative to a second beam.
In the present experiment two beams are bent, one in an electrostatic guide and one in a magnetic guide. 

$\mathrm{Ne}^*$+\nhds is studied by using two separate pulsed (20 Hz) supersonic expansions, SE1 and SE2 in figure \ref{expt}. 
The entire setup is housed in a fourfold differentially pumped high-vacuum setup.
The four differentials are a source chamber, guide chamber, collision chamber, and the last chamber that contains detectors for beam density normalisation.
The base pressure in the guide and collision chambers is in the mid-10$^{-8}$ mbar range.
It rises to about $2\cdot 10^{-7}$ mbar in the guide chamber and remains unchanged in the collision chamber.
The source chamber contains both supersonic expansions, separated by a metal wall above which a single turbo pump is mounted.
Both guides are in the same vacuum chamber which is pumped by two turbo pumps mounted at either end of the 1.5 m long, kinked chamber.

$\mathrm{Ne}^*$, formed in SE1, is guided in a magnetic guide (MG in fig \ref{expt}). 
SE1 is a liquid-nitrogen cooled Even-Lavie valve \cite{Even:2000eg} where the $\mathrm{Ne}^*$ atoms are produced by electron bombardment directly behind the source. 
Neat neon is used at a backing pressure of 10-20 bar, and the speed of the beam is controlled by adjusting the valve temperature in the range 200-320 K. 
This produces velocities in the range 530-850 m/s. 
The beam is skimmed and enters the MG in the second differential high-vacuum chamber. 
The MG is built from NdFeB permanent magnets with a remanent field of 1.17 T. 
It is composed of a straight section with hexapole symmetry and a curved section with quadrupole symmetry. 
The magnetic field distributions in the two sections are shown in figures \ref{expt}(b) and (c), respectively.
In the first section the transverse potential is approximately 0.7 T deep, while in the second section it is 0.5 T deep towards the outside of the bend. 
Using the configuration shown in figure \ref{expt}(c) leads to reduced confinement of the paramagnetic particles, but it is required in order to provide side-access for the beam of polar molecules emerging from the electrostatic guide. 
The guiding force on the $\mathrm{Ne}^*$ atoms is produced by the Zeeman effect and the inhomogeneous magnetic fields in the guide. 
For the $^3P_J$  state the first order Zeeman effect is given as 
\begin{equation}
W_Z=g_{\alpha J}\mu_B BM,\label{eq:zeeman}
\end{equation}
where $\mu_B$ is the Bohr magneton, $B$ is the magnetic field magnitude, and $M$ is the magnetic quantum number that can take integer values $-J<M<J$.
$g_{\alpha J}$ is the gyromagnetic ratio and is given by
\begin{equation}
g_{\alpha J}=1+\frac{J(J+1)+S(S+1)-L(L+1)}{2J(J+1)}.\label{eq:gyro}
\end{equation}
Equations \ref{eq:zeeman} and \ref{eq:gyro} show that of the three spin-orbit components of the $^3P_J$  state only J=1 and 2 can be guided. 
Of these two only J=2 has a sufficient lifetime because the J=1 component quickly decays to the ground state. 
The bent guide, having a radius of curvature of close to 6 m obtained by joining several straight segments with low angles between them to form an overall bend of 10 degrees, eliminates all diamagnetic states from the beam.
As a consequence, the $\mathrm{Ne}^*$ atoms that reach the interaction region are only $^3P_2$. 
Trajectory calculations using simulated magnetic field distributions predict an acceptance in longitudinal velocity of the magnetic guide for Ne$^*$ up to 800 m/s. 

\begin{figure}
 \includegraphics[width=\figwidth]{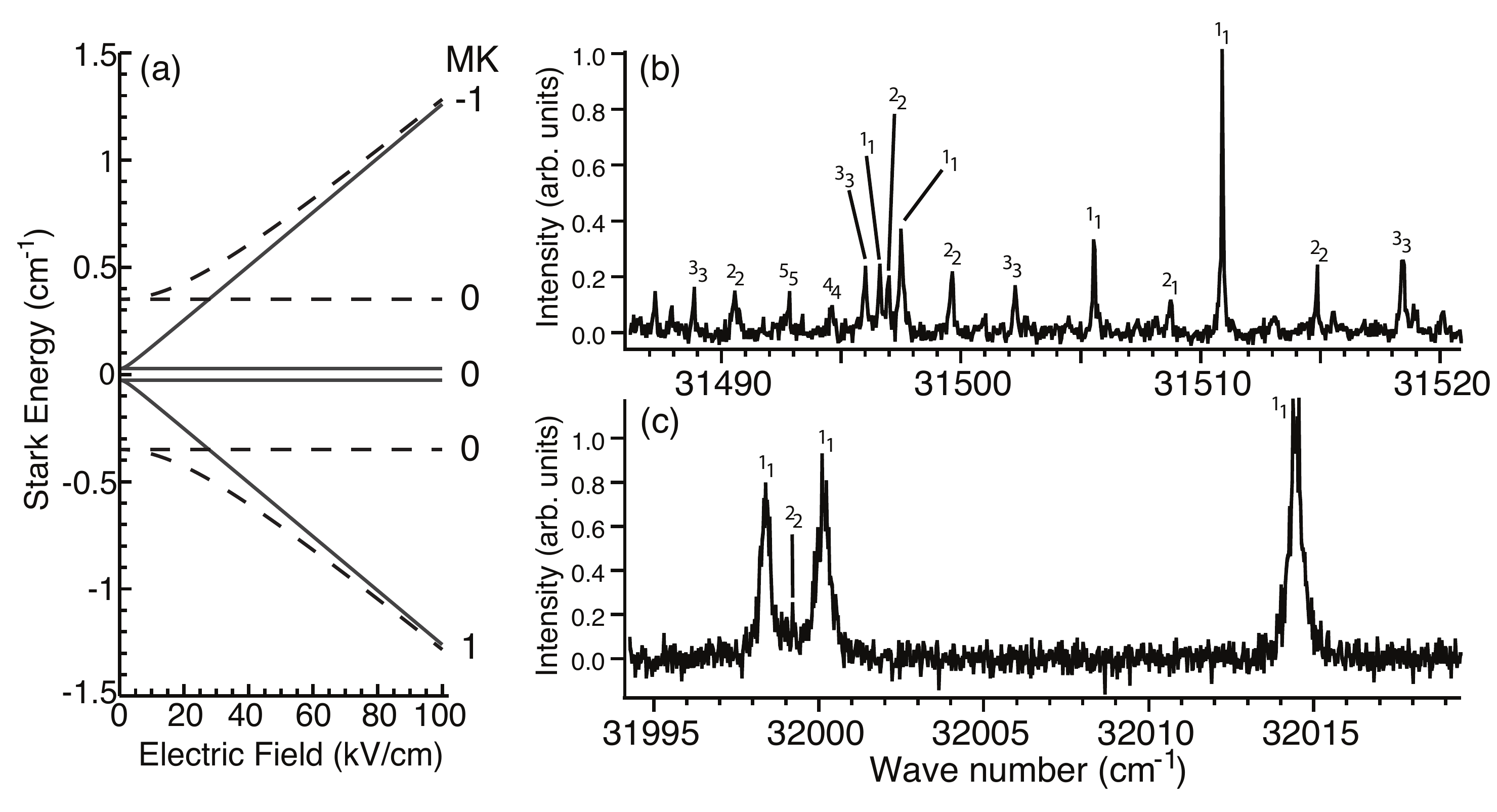}%
\caption{\label{states}(a) Stark Effect for the \jk{J}{K}=\jk{1}{1} for \ndds (solid lines) and \nhds (dashed lines). (b) REMPI spectrum of \ndds recorded in TOF2 (see figure \ref{expt}) for a mixture of 8\% \ndds in neon. (c) REMPI spectrum of \nhds recorded in TOF2 for a mixture of 8\% \nhds in neon. Labels above each transition designate the initial state as \jk{J}{K}.}
 \end{figure}

The ammonia beam line starts with a room-temperature general valve (Parker Series 9), SE2, where \nhds is expanded from a backing pressure of 1.5 bar either pure (beam velocity 1080 m/s) or seeded in neon (beam velocity 830 m/s) or argon (beam velocity 600 m/s) at a 1:10 ratio. 
The beam is skimmed and enters the electrostatic guide (EG in fig \ref{expt}) in the second high-vacuum chamber. 
The EG is a hexapole guide built from straight segments that are each 10 cm long. 
Consecutive segments subtend an angle of one degree, leading to a similar bend angle as in the MG. 
The radius of curvature is also the same as that of the MG. 
The cross section through the EG, plotted in figure \ref{expt}(d), shows the electric field distribution on the inside of the guide. 
Application of $\pm$11 kV, as generally used in the present study, produces an electric field maximum between two electrodes of $\approx$100 kV/cm. 
This permits to guide \nhds molecules with forward velocities up to $\approx$1000 m/s and \ndds up to $\approx$1100 m/s. 
The guiding of the heavier isotopologue is facilitated by the small inversion splitting (0.079 cm$^{-1}$ vs. 0.5 cm$^{-1}$) which leads to a linear Stark effect at substantially lower electric fields (see figure \ref{states}(a)). 
Conditions in the supersonic expansion presumably lead to formation of large quantities of ammonia clusters, but these are eliminated in the electrostatic guide since they are either non-polar or become high-field seeking at moderate electric fields already. 
The last component of the EG is a 30 cm long electrostatic hexapole lens that is used to focus the polar molecules through the magnets of the bent section of the MG and merge the two beams. 
The lens is operated at voltages of up to $\pm$3 kV and optimised to obtain maximum overlap between the two beams. 

Figure \ref{states}(a) shows the Stark effect for the \jk{J}{K} = \jk{1}{1} states of \nhds (dashed lines) and \ndds (solid lines). In each case the zero-field splitting produces two states, normally called $\tilde{X}(1)$ (the upper component of the doublet) and $\tilde{X}(0)$ (the lower component). 
In the presence of an electric field all states from $\tilde{X}(1)$ are shifted to higher energies and all states from $\tilde{X}(0)$ to lower energies. 
A static inhomogenous electric field with a minimum in two or three dimensions thus allows to confine only the low-field seeking $\tilde{X}(1)$ states while the $\tilde{X}(0)$ are pushed away. 
The force that can be exerted on the molecules depends on the Stark effect which in turn depends on the quantum numbers J, K, as well as on the projection $M_J$ of J on the field axis.\cite{vandeMeerakker:2012ft}
We have previously studied, using resonance-enhanced multi-photon ionization (REMPI), the guiding probability of different (J,K,M) states in bent electrostatic hexapole guides and found that while such devices indeed purge a molecular beam from all the high-field seeking states, they do not substantially alter the distribution of \jk{J}{K} states in the remaining $\tilde{X}(1)$ state.\cite{Bertsche:2011jr,Bertsche:2010vr}
In the present experiment we employed the same REMPI approach to characterize the rotational temperature of the guided beam. 
Figures \ref{states}(b) and (c) show (2+1) REMPI spectra, using the well-investigated two-photon $B(v_2'=5)\longleftarrow X(v_2''=0)$ transition,\cite{Allen:1991tb} recorded 80 cm behind the end of the EG. 
Transitions are labeled by the quantum numbers \jk{J}{K}, and it is immediately evident that all states with K=0 are missing. 
These states have no appreciable Stark effect and are not guided. 
By comparing the experimental spectra in Fig. \ref{states} with calculated spectra using the pgopher programme\cite{Anonymous:v7enFBFX} we conclude that when using a pure expansion of \nhd (\ndd) $\approx$94\% ($\approx$79\%) of the total guidable flux are in the lowest three guidable rotational states, \jk{J}{K}=\jk{1}{1}, \jk{2}{1} and \jk{2}{2}. 
When seeding ammonia in neon we find $>$99\% and $>$96\% in these states for \nhds and \ndd, respectively. 
The population in the \jk{1}{1} state for pure expansions is 55\% and 32\%, and for seeded expansions 86\% and 58\% for \nhds and \ndd, respecivelty. 
No attempt was made to measure the population in the vibrationally excited states but given the large vibrational frequencies in ammonia we can assume that even in a room-temperature expansion these populations can be neglected.

30 cm downstream from the end of the MG the reaction products are detected, using a pulsed Wiley-McLaren type mass spectrometer, labeled TOF1 in figure \ref{expt}.\cite{Wiley:2004bl}
Since one of the products of the present reaction is charged it can be detected without further effort. 
TOF1 is switched on, by the application of a 300 ns, 1000 V pulse to the extraction plate, at a specific time that depends on the particular velocities that are recorded. 
The flight tube of TOF1 is always at high voltage and all ions that are formed outside the extraction region are deflected away from the beam axis.
Only ions that are formed inside TOF1 during the application of the extraction pulse are accelerated onto a microchannel plate detector (MCP). 
This way a narrow slice from each of the pulses is selected and used for the collision study.
The resolution of TOF1 is sufficient to separate fragments with 17 and 18 atomic mass units. 
By counting product ions during a specified amount of time we measure relative reaction rates for different collision energies. 
In order to convert these to rate constants and reaction cross sections (both on arbitrary scales) we use the relations 
\begin{equation}
\Delta[NH_3^+]/\Delta t=\mathrm{k}(v_{rel})N_{NH_3} (t)N_{\mathrm{Ne}^*}(t)\label{eq:rate}
\end{equation}
and $k(v_{rel}) = v_{rel}\sigma(v_{rel})$, where $k(v_{rel})$ is the rate constant at relative velocity $v_{rel}$, $\sigma$ is the cross section, and the $N$ are the reactant densities which are assumed to be constant throughout each experimental cycle.
This assumption is reasonable in view of the relative durations of extraction (300 ns) and molecular pulse (100s of $\mu$s in the interaction region).
It is also supported by the measured, total Ne* density that is independent of the presence of the ammonia beam.
In order to convert the observed count rates to rate constants and cross sections it is imperative to have an accurate (even if relative) calibration of the beam densities. 
To this end we have mounted two detectors 50 cm downstream from TOF1 where the ammonia and $\mathrm{Ne}^*$ densities are monitored separately. 
TOF2 in figure \ref{expt} is used to measure the ammonia density, using the same REMPI process described above. 
\nhdps ions formed by ionization from the \jk{J}{K}=\jk{1}{1} state are accelerated perpendicularly to the beam axis and recorded at every cycle of the experiment. 
Simultaneous measurement of the laser power and appropriate scaling of the \nhdps signal then permits to calculate a relative density of the ammonia beam at every shot, and average it for the entire duration of an experiment. 
$\mathrm{Ne}^*$ is measured directly by impinging the beam on an MCP placed on-axis with the neon beam (D in figure \ref{expt}). 
Since the molecular pulses can vary both in overall intensity and in shape it is important to probe the density precisely in that part of the pulse that is also used for the collision. 
Both the timing of the laser for REMPI and the time to read out the $\mathrm{Ne}^*$ intensity are chosen accordingly.
To measure the rate constant for a particular collision energy the two beam velocities have to be set appropriately. 
The ammonia expansion is chosen either pure or seeded, and the temperature of SE1 is set to obtain the desired relative velocity. 
The speed ratios in both expansions are non-zero, and the desired collision energy is obtained only if the relative timings are set to particular values. 
The velocities then define the precise timing of the extraction pulse in TOF1, as well as the timings for the beam monitoring in TOF2 and on D. 

\begin{figure}
 \includegraphics[width=\figwidth]{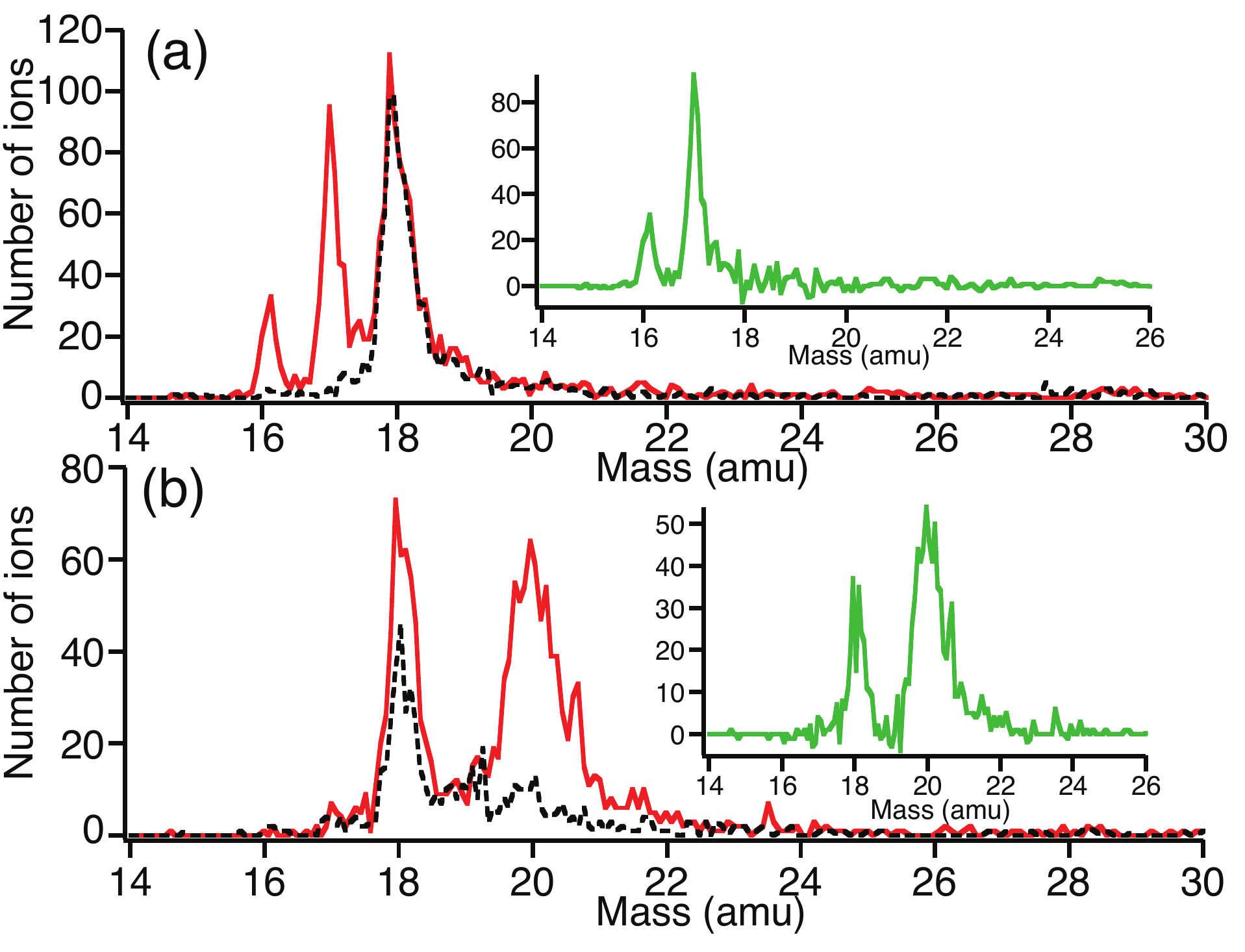}%
\caption{\label{MS}Mass spectra recorded for the detection of reaction products in TOF1 (see figure \ref{expt}). Results for (a) \nhd, and (b) \ndd. In each case the red trace shows the data for perfect valve timings and the black trace shows a background scan recorded with the ammonia beam delayed by a ms. The green traces in the inset show the difference spectra that are the actual number of reaction products for both isotopologues.}
 \end{figure}
The low reaction rates require the accumulation of product ion signals over thousands of experimental cycles. 
A single experimental run lasts between 5 min and 30 min, corresponding to up to 40000 pulses. Ions that reach the MCP in TOF1 are counted at their particular arrival time to preserve the mass information. 
In case of neat ammonia and 8\% \nhd/Ne gas mixtures, up to 1000 \nhdps product ions are collected in 5 minutes at the highest collision energies studied. 
A considerably slower 8\% \nhd/Ar mix yields as few as 100 \nhdps ions in 15 minutes at the lowest collision energies. 
A typical trace that results from such a measurement is shown in figure \ref{MS} for \nhds (panel (a)) and \ndds (panel (b)). 
The full trace in panel (a) shows three prominent peaks at masses 16, 17, and 18 that correspond to \nhdp, \nhtp, and H$_2$O.
The H$_2$O originates from background water that is also Penning ionised by the Ne* atoms.
Only masses 18 and 20 are present in panel (b).
Here, the heavier ions are purely \nddps but the peak at 18 amu is composed of both \ndtps and H$_2$O. 
In order to get the pure \ndtps signal a background trace is recorded under identical conditions except that the ammonia pulse is delayed by 1 ms. 
The two traces, shown as red solid and black dashed lines in the main panels of fig \ref{MS} are then subtracted to obtain the pure signals, shown in the inset, for this particular collision energy. 
The rate constant is obtained by integrating each of the peaks and dividing by the normalization signals obtained in TOF2 and D.

The resolution in the collision energy in our experiment is determined by the range of velocities that contribute to any single data point.
The latter is determined by the speed ratios $v/\Delta v_{expansion}$ and pulse durations of both expansions. 
As described by Narevicius et al.\cite{Shagam:2013ev} a flight time that is long relative to the pulse duration leads to a rotation of the phase-space distribution of the pulse which in turn leads to an energy resolution in the experiment that is higher than what is given by the speed ratio itself. 
The experimental resolution is further improved by extracting the reaction products in a pulsed TOF that enables us to cut a narrow slice out of the total velocity distribution.
We estimate the resolution in our experiment by calculating the highest and lowest velocities $v_{max}$ and $v_{min}$ that molecules can have to reach TOF1 during the application of the extraction pulse. 
Using $\tau$ as the flight time of the molecules through the guide, $\Delta T$ for the duration of the molecular pulse, and $\Delta t$ for the duration of the extraction pulse we get
\begin{equation}
\tau_{max;min}=T_0\pm \frac{\Delta T}{2}-(t_0\mp \frac{\Delta t}{2}),
\end{equation}
where the upper signs give ${{\tau }_{max}}$ and the lower signs give ${{\tau }_{min}}$.  
The limits of $\tau $ are then used to calculate the maximum possible $\Delta v$:
\begin{equation}
\Delta v_{TOF}=v_{max}-v_{min}=\frac{L}{\tau _{min}}-\frac{L}{\tau_{max}},
\end{equation}
where $L$ is the length of the guide. 
To determine the actual velocity spread in the collision experiment this value has to be compared with the spreads in the expansions, $\Delta v_{expansion}$, and the actual spread then is $\Delta v=\text{min}\{\Delta v_{TOF},\Delta v_{expansion}\}$. 
The experimental values for the opening times are 35 $\mu$s for SE1 and 60 $\mu$s for SE2, and the flight distance is $\approx $2 m, resulting in $\Delta v=$ 30 m/s at $v$=1000 m/s and 20 m/s at 800 m/s. 
The measured speed ratios (around 5-20 for both expansions) give a larger spread. 
Since $T_{0}=L/v_{0}$, $\Delta v$ does depend slightly on the absolute beam velocities.

Assuming full control over the longitudinal velocity distribution and elimination of its spread the ultimate limitation of the experimental resolution in a merged beams experiment would be given by the transverse velocity spread.
In our setup there are three factors that may determine the transverse spread: the geometric selection of the components of the beam that are fed into the guide, the acceptance of the guides themselves, and the geometric selection between the end of the guide and the interaction region.
The transverse acceptances of both guides are on the order of 10-20 m/s for the present particles, and the emittance of the source (including the skimmer) is matched to that.
At the end of the guide the geometric selection is given by the radii of guides and interaction region, and the distance between the two.
Having guide radii of 8 mm and a radius of the interaction region of 5 mm we can use the distance of 300 mm to calculate a ratio of transverse-to-longitudinal velocities of 40-100.
At the lowest relative longitudinal velocity this would correspond to transverse velocities up to 10 m/s. 
The same value is also obtained in trajectory calculations where we obtain a distribution of $\pm$ 5 m/s.
It has to be noted that the actual distribution in relative transverse velocities is smaller than that because the particles with the extreme transverse velocities in each of the beams will not cross inside the interaction region.
Consequently the transverse relative velocity distribution at this point is at least a factor ten smaller than the longitudinal spread and is neglected in the data interpretation.

\section{THEORY}

\begin{figure}
 \includegraphics[width=\figwidth]{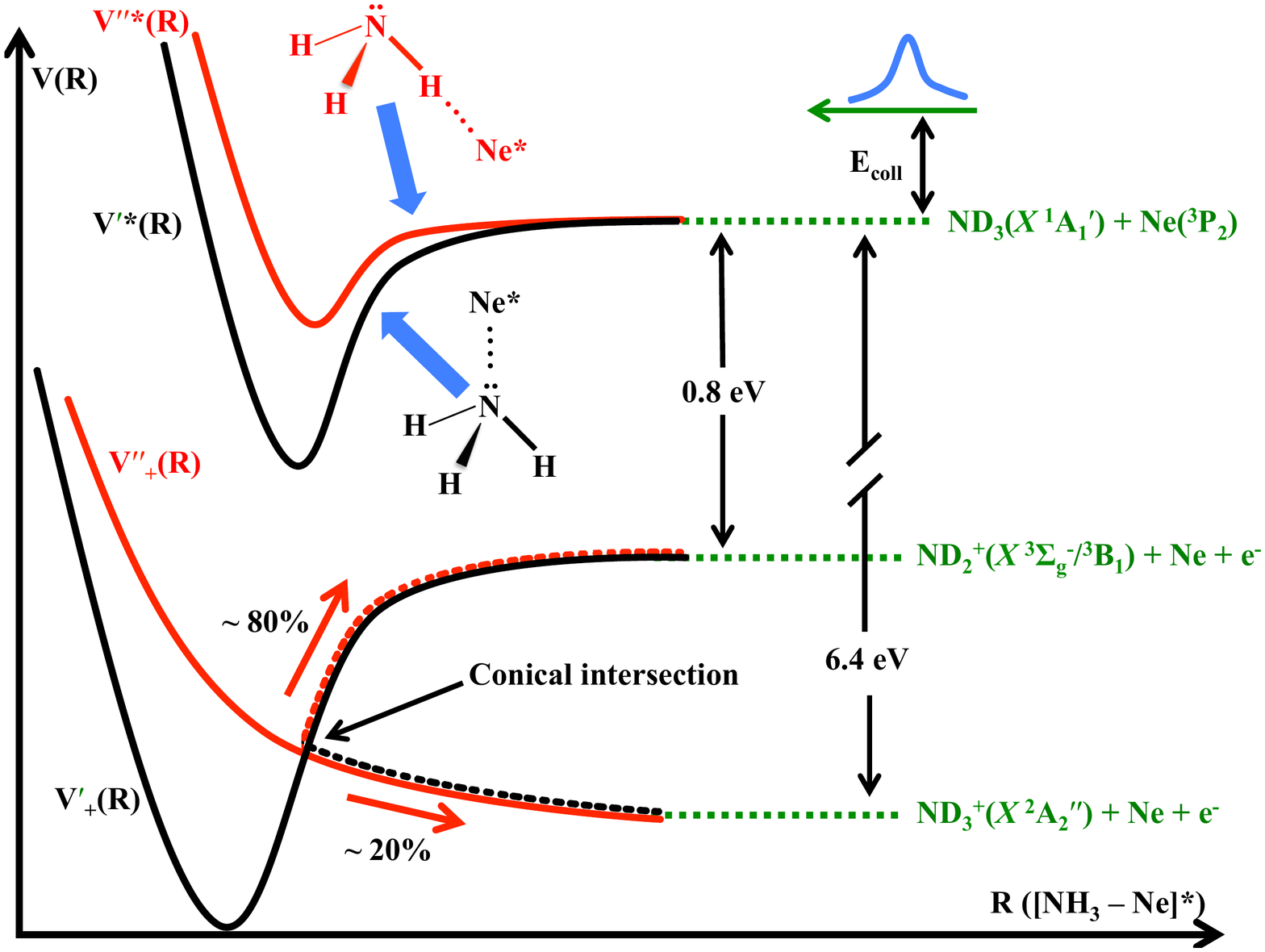}%
\caption{\label{potentials}Schematic view of the potential curves which provides an intuitive explanation of the Ne*+\nhds reaction mechanism. 
The curves labelled by V* show the excited states (Ne*+\nhd) with the top red one corresponding to the neon atom approaching ammonia along one of the N--H bonds and the black one to the approach along the lone pair axis. 
The V$_+$ potentials are the ionic curves on which the system lands after the PI. The red ionic curve corresponds to the excited $\tilde{\mathrm{A}}$ $^2$E state of \nhdp, the black curve is the ground state. 
Molecules in the $\tilde{\mathrm{A}}$ state can dissociate, or they can relax to the ground state via a conical intersection.}
 \end{figure}
 
A simple schematic view of the present reaction can be given by the optical potential model, illustrated by figure~\ref{potentials}.\cite{Niehaus:1981vb,Siska:1993cf}
The model operates only on four potential curves; the excited A*+B curves, corresponding to two different electronic orbitals which may take part in the process, the N--H bond or the lone pair centered on nitrogen atom, and the ionic A+B$^+$ curves describing the reaction products. 
Coupling between the initial and final states can be modelled by the complex part of the V* potential
\begin{equation}
V(R)=V^*+i\Gamma(R)/2.
\end{equation}
Ben-Arfa et al. have argued that the approach along the lone pair favors  electron transfer from the \nhd (2a$_1$)$^2$ (lone pair) molecular orbital into the Ne* core, with a simultaneous/subsequent autoionization of neon.\cite{BenArfa:1999el}
This produces an \nhdps ion in the ground electronic state. 
Conversely, a reaction that passes through the $\mathrm{Ne}^*$--H--N configuration results in a ground state neon atom and an \nhdps ion in an electronically excited state, because the electron has been lost from one of the \nhd (1a$_1$)$^2$(1e)$^4$ bond orbitals. 
It has been shown that over 80\% of the \nhdp($\tilde{\mathrm{A}}$ $^2$E) molecular ions dissociate into \nhtps and H products.  
More importantly, it has been suggested that the collision energy has little effect on the fraction of \nhdp($\tilde{\mathrm{A}}$ $^2$E) that dissociates.  
Simultaneous detection of \nhdps and \nhtps reaction products is therefore a stereodynamic probe of the $\mathrm{Ne}^*$ + \nhds molecular collisions.      

Due to the spin-orbit coupling the full {\it ab initio} description of collisions in the Ne($^3$P)+NH$_3$ complex involves dynamics taking place on nine coupled potential energy surfaces (PESs).
The interaction energy of Ne* and NH$_3$ has been found to be similar to that in the Rb-NH$_3$ complex (see below), namely several 100 meV\cite{Pzuch:08}, the spin-orbit coupling for Ne($^3P$) is small in comparison ($\approx$ 50 meV).
As a consequence, all $J$ and $M$ components of the $^3P$ state need to be considered. 

A straightforward treatment of such a complicated system, including nonadiabatic couplings between PESs, their avoided crossings and conical intersections, currently is not feasible even with present multireference {\it ab initio} methods. 
Nevertheless, one can refer to simplified models in order to exctract information necessary for scattering calculations. 
Recent examples where this has been done include the NO($X ^2\Pi$)--OH($X ^2\Pi$) and OH($X ^2\Pi$)--ND$_3$ systems, for which electronic structure calculations were limited either to a multipole expansion model, or two lowest adiabatic PESs.~\cite{Kirste:12,Sawyer:11}

For systems undergoing Penning Ionization an additional challenge arises. 
Due to the coupling of the A* + B complex with a continuum of states of the (AB)$^+$ + $e^-$ type, application of the standard {\it ab initio} methods based on a variational principle, such as multi reference configuration interaction (MRCI) commonly used in multi-surface cases, may result in driving down the initial state either to the ground state, one of the excited states or to some delocalized state corresponding to the fragmentation into an ion, a molecule and a free electron. 
It is to be noted that Penning Ionization with molecules such as ND$_3$ has never been characterized by a full {\it ab initio} approach, even for simpler systems involving He($^3$S). 

To obtain interaction energies that are unambigously related to the specified monomers, one may use the SAPT(UHF) formalism. Due to its perturbative nature, it would not permit for any variational collapse of the dimer or any unwanted modifications of the monomers. Therefore, SAPT(UHF) is well suited for treatment of the Penning Ionization process.

Indeed, an open-shell formulation of SAPT has recently been applied for systems involving excited He($^3$S) and Ne($^3$P) atoms.~\cite{Hapka:12,Hapka:13}  
In the present work SAPT(UHF) is used to construct a model state-averaged potential and to obtain the leading long-range coefficients for Ne*--NH$_3$. 
However, since application of SAPT to excited states involves convergence problems, the description of the dispersion interaction is given only in the asymptotic region of the potential energy surface.
The reaction process itself will then be described by means of quantum defect theory.

The geometry of the Ne*-ND$_3$ complex is given in Jacobian coordinates, where $r$ is the distance from the center of the mass of ND$_3$ to the atom, $\theta$ stands for the angle between the intermolecular vector and the $C_3$ axis of the NH$_3$ molecule ($\theta$ = 0 corresponds to the atom approaching towards the lone pair of the molecule) and $\phi$ denotes the dihedral angle between the plane containing the $C_3$ axis and an NH bond and that containing the $C_3$ axis and the intermolecular vector. 
The geometry of ND$_3$ is frozen in the $C_{3v}$ symmetry with N-D bond lenghts of 1.913 $a_0$ and N-H-N angles equal to 106.7$^{\circ}$.~\cite{Pzuch:08,Benedict:57} 
All SAPT calculations are performed in the developer's version of the MOLPRO package.~\cite{molpro}
    
The $V_{\rm disp}(r,\theta,\phi)$ interaction potential was calculated on a grid with intermolecular distances $r$ from 30 to 150 $a_0$ ($a_0$ is the Bohr radius) and two azimuthal angles $\phi$ = 0 and 60$^{\circ}$. The $\theta_i$ angles were chosen to be the points for 5-point Gauss-Lobatto quadrature. In order to reduce the computational effort the $\phi$ dependence in the $V_{\rm disp}(r,\theta,\phi)$ potential was represented in the form
\begin{equation}
V_{\rm disp}(r,\theta,\phi) = \sum_{k=0}^{\infty} V_{3k}(r, \theta) \cos 3k\phi
\end{equation}
and the leading terms, $V_{0,{\rm disp}}(r, \theta)$ and $V_{3,{\rm disp}}(r, \theta)$, were approximated by the sum and difference potentials
\begin{equation}
 \begin{aligned}
 V_{0,{\rm disp}}(r, \theta) &= \frac{1}{2}\left[V_{\rm disp}(r,\theta,0^{\circ}) + V_{\rm disp}(r,\theta,60^{\circ}) \right] \\
 V_{3,{\rm disp}}(r, \theta) &= \frac{1}{2}\left[V_{\rm disp}(r,\theta,0^{\circ}) - V_{\rm disp}(r,\theta,60^{\circ}) \right]
       \end{aligned}
 \qquad 
\end{equation}
The d-aug-cc-pVTZ basis set was selected for the neon atom, aug-cc-pVTZ for hydrogen and nitrogen atoms.~\cite{Dunning:89,Kendall:92} 
Because of the expected small anisotropy for Ne($^3$P) complexes\cite{Hapka:13,Gregor:81} we composed a state-averaged potential.

The leading dispersion coefficients, $C^{00}_{6,{\rm disp}}$ and  $C^{10}_{7,{\rm disp}}$, were obtained by expanding $V_{\rm disp}(r, \theta, \phi)$ in renormalized spherical harmonics according to Eq. 1 of Ref. \onlinecite{Pzuch:09}, and fitting the corresponding potential expansion coefficients, $V_{00}(r)$ and $V_{10}(r)$. 
The $C^{00}_{6,{\rm disp}}$ and $C^{10}_{7,{\rm disp}}$ in atomic units are 254.4 and -61.5, respectively. 
The dipole and quadrupole moments of ND$_3$, as well as the state-averaged polarizability of Ne($^3$P), necessary for getting the long-range induction coefficients were calculated with the finite field method at the CCSD(T)/aug-cc-pVQZ level of theory, giving 0.599 $ea_0$, -2.0258 $ea_0^2$, and 183.44 $a_0^3$, respectively. 
In order to obtain the leading van der Waals induction coefficient $C^{00}_{6,{\rm ind}}$, suitable asymptotic expressions of the multipole expansion were used.~\cite{Schmuttenmaer:91}.
The obtained induction coefficients equal $C^{00}_{6,{\rm ind}}$ = $C^{20}_{6,{\rm ind}}$  = 65.8, $C^{10}_{7,{\rm ind}}$ = -801.4 and $C^{30}_{7,{\rm ind}}$ = -534.3.

\begin{figure}
\includegraphics[width=\figwidth]{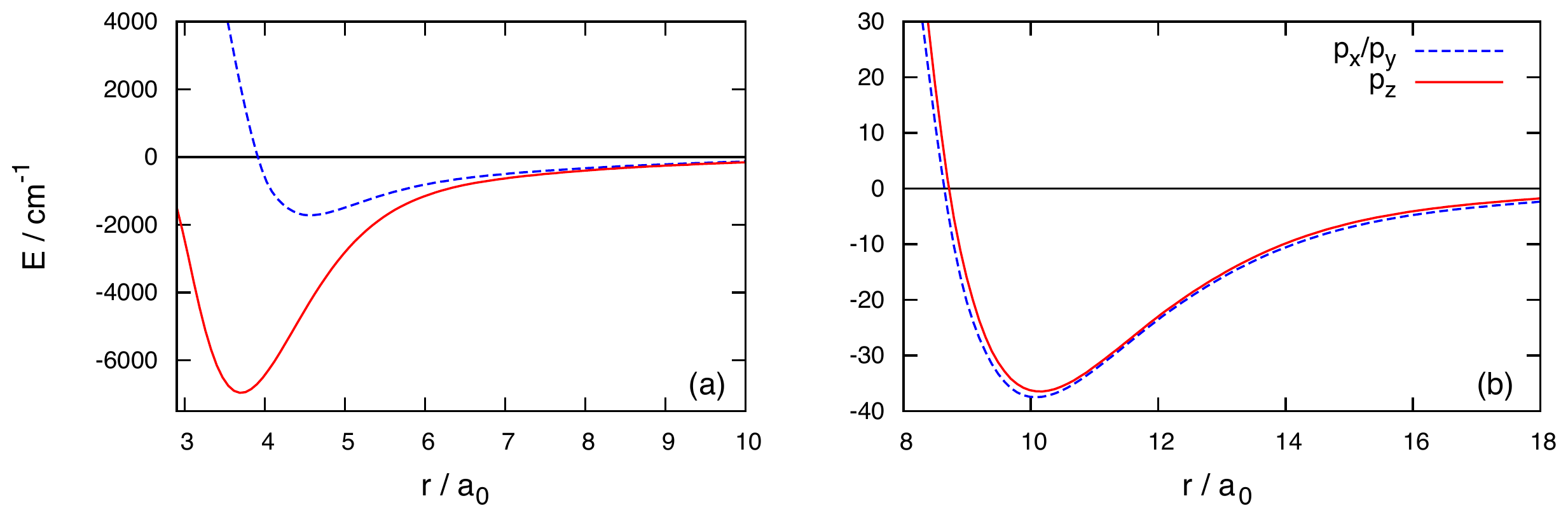}
\caption{\label{minima}Interaction potential for the Ne*-ND$_3$ complex in colinear geometries. Panel a) corresponds to the Ne*-ND$_3$ ($\theta$ = 0$^{\circ}$, $\phi$ = 0$^{\circ}$) and panel b) to the Ne*-D-N ($\theta$ = 180$^{\circ}$, $\phi$ = 0$^{\circ}$) geometry. $p_{x,y,z}$ refers to the unpaired $2p$ electron of the Ne* atom.}
 \end{figure}

For the collinear geometry of the Ne*-ND$_3$ complex we also performed SAPT(UHF) calculations up to the second order in intermolecular interaction operator taking into account three spatially degenerate states of the Ne* atom. In the case of the Ne*-lone pair-N geometry ($\theta$ = 0$^{\circ}$, $\phi$ = 0$^{\circ}$) we found a minimum of 6961 cm$^{-1}$ located at 3.7 bohr corresponding to the electron being located at the $p_z$ orbital of Ne ($\langle L_z^2 \rangle$ = 0, see Fig.\ref{minima}). Two remaining states in this geometry give a minimum of 1713 cm$^{-1}$ at 4.6 bohr.

For the Ne*-D-N geometry ($\theta$ = 180$^{\circ}$, $\phi$ = 0$^{\circ}$) minima occur at 10 bohr D$_e$ = 37.5 cm$^{-1}$ and at 10.2 bohr, $D_e$ = 36.4 cm$^{-1}$, for $\langle L_z^2 \rangle$ = 1 and $\langle L_z^2 \rangle$ = 0 states of the Ne* atom, respectively (Fig. \ref{minima}). 

As expected, the minima occuring at large distance for the Ne*-D-N arrangement exhibit small anisotropy due to the effective screening of the inner $2p$ shell of the neon atom by the outer $3s$ electron. In contrast, when Ne* approaches the lone pair localized at the nitrogen atom, the strong induction and dispersion attraction leads to minima at much shorter distace resulting in alleviated screening and strongly pronounced anisotropy. Finally, it is important to notice that the obtained interaction potentials are similar in character to the ones observed for NH$_3$ alkali-metal atoms, e.g. Rb-NH$_3$ or Li-NH$_3$.~\cite{Pzuch:08}

Apart from the aforementioned terms, the quadrupole-dipole interaction decaying as $r^{-4}$ needs to be discussed. 
This term, however, does not significantly affect the reaction dynamics in the energy range studied here.
Neglecting this term in modelling the Ne*-ND$_3$ collisions can be rationalized by assessing the range at which the Ne*-ND$_3$ interaction becomes dominated by the electrostatic contribution. 
Such an estimation has been conducted by comparing the electrostatic and second order (i.e. dispersion and induction) SAPT energy contributions for collinear geometry of the complex in $A_1$ symmetry of the $C_{3v}$ point group, which corresponds to the electron localized on the $p_z$ orbital of Ne*. 
The findings suggest that the electrostatic energy starts to dominate at distances larger than 30 $a_0$. 
It can be explained by the small value of the Ne($^3$P) quadrupole moment, which at the MRCISD level of theory amounts to 0.245 $ea_0^2$. 
Given the lowest energies available in the experiment, i.e. 0.1 K, and the number of partial waves included in scattering calculation described in the following paragraphs one can establish the position of the centrifugal barrier for Ne*-ND$_3$ at roughly 26 $a_0$. 
Therefore the main contribution to the potential in the region of interest for our reaction model should originate from the dispersion and induction energy contributions, an effect captured by the model potential employed here. 
The terms proportional to $r^{-4}$ terms would be important only at lower energies.

The knowledge of the asymptotic region of the potential energy surface provided by \textit{ab initio} calculations allows the use of multichannel quantum defect theory (MQDT) to obtain the reaction rates. 
MQDT takes advantage of the fact that the chemical reaction takes place at relatively short distances compared to characteristic van der Waals length. 
It will be shown how this observation can be used to significantly simplify the description. 
The following is a review of the general MQDT formalism, using the notation introduced by Mies.\cite{Mies1984a,Mies1984b} 
The total wave function can be decomposed into angular and radial part
\begin{equation}
\boldsymbol{\Psi}=\sum_{i=1}^{N}{\frac{1}{r}\Phi_i(\theta,\phi)\psi_i(r)}.
\end{equation}
It has to satisfy the multichannel Schr\"{o}dinger equation

\begin{equation}
\sum_j{\left[\left(-\frac{\hbar^{2}}{2\mu}\frac{d^{2}}{d\,r^{2}}+\frac{\hbar^{2}\ell_i(\ell_i+1)}{2\mu r^{2}}+E^{th}_i\right)\delta_{ij}+W_{ij}\right]\psi_j(r)}=E\psi_i(r) \,,
\end{equation}
where $r$, $E$, $\mu$, $E^{th}_i$ and $\ell_i$ are the intermolecular distance, energy, reduced mass, threshold energy for \textit{i}th channel and its orbital angular momentum, respectively, and the $W_{ij}$ are the interaction matrix elements, which include the long-range (diagonal) interactions as well as the interchannel couplings and short-range potential details.
In the present case three kinds of channels corresponding to the initial states and two possible reaction products are considered, with a large number of partial waves in each channel.
All other channels are neglected, in particular any rovibrational excitations and other electronic states. 
In this case the kinetic energy is higher than all the threshold energies, so all the channels are open. 
The form of the interaction at large distances in the entrance channel can be found using SAPT, as discussed in previous paragraphs. 
However, at short distances the form of interaction can be very different. 
One also has to include couplings between the partial waves resulting from anisotropies in the PES.
  
In MQDT the true interaction $\mathbf{W}$ is replaced by a set of reference potentials which reproduce the asymptotic form of $\mathbf{W}$ at large distances. 
In addition, a quantum defect matrix $\mathbf{Y}$ is introduced to parametrize the short range effects. 
For many problems, including the present one, the channels can be regarded as uncoupled beyond some distance, and an intuitive choice for the reference potentials is to take only the diagonal part of the interaction. 
The solution of the new problem at short distances can be expressed using wave functions $\hat{f}$, $\hat{g}$ with WKB-like form. 
The $\mathbf{Y}$ matrix connects these functions with the long-distance solutions of the scattering problem $f$, $g$ using the MQDT functions $C(E)$ and $\tan\lambda(E)$ (when closed channels are present, an additional function $\nu(E)$ is required)~\cite{Julienne1989,Mies2000}
\begin{align}
\label{functions}
\begin{array}{lll}
f_i(r)    & = & C_i^{-1}(E)\hat{f}_i(r)\\
g_i(r)    & = & C_i(E)(\hat{g}_i(r)+\tan\lambda_i\hat{f}_i(r)).
\end{array}
\end{align}
The general solution of the coupled channels problem can be written as
\begin{equation}
\mathbf{F}(r)=(\mathbf{f}^0(r)+\mathbf{Y}\mathbf{g}^0(r))\mathbf{A},
\end{equation}
where $\mathbf{f}^0(r)$ and $\mathbf{g}^0(r)$ are diagonal matrices of the reference solutions and $\mathbf{A}$ denotes the amplitudes. 
The scattering matrix $\mathbf{S}$ is given by means of the $\mathbf{Y}$ matrix and the diagonal matrices, containing MQDT functions and phase shifts~\cite{Mies1984a}
\begin{equation}
\label{SMQDT}
\mathbf{S}=e^{i\boldsymbol{\xi}}(1+i\mathbf{R})(1-i\mathbf{R})^{-1}e^{i\boldsymbol{\xi}},
\end{equation}
where
\begin{equation}
\label{RMQDT}
\mathbf{R}=\mathbf{C}^{-1}(E)(\mathbf{Y}^{-1}-\tan\boldsymbol{\lambda}(E))^{-1}\mathbf{C}^{-1}(E)
\end{equation}
and $\xi_i$ is the phase shift in \textit{i}th channel induced by the chosen reference potential $V_i$. 
Equations~\eqref{SMQDT}-\eqref{RMQDT} are particularly convenient when the long range interaction is dominated by a single $1/r^{4}$ or $1/r^{6}$ term which allows the use of analytical solutions of the Schr\"{o}dinger equation.~\cite{Gao1998a,NJP2011} 
In numerical calculations the wave function usually is propagated using MQDT boundary conditions given by eq. ~\eqref{functions}, and the results are compared at very large distances with the asymptotic solutions to find $\mathbf{S}$.

One of the benefits of this formulation of the problem is that if the long range interactions are correctly accounted for, the $\mathbf{Y}$ matrix describes only the short range processes for which the energy scales are usually significantly larger than energies characteristic for long range interactions. 
As a result, the quantum defect matrix only weakly depends on the kinetic energy and orbital angular momentum.

In the present reaction an important simplification to the problem comes from the fact that the exit channels have threshold energies far below the entrance channel.
Indeed, the difference between exit and entrance channels is more than 1 eV which is over six orders of magnitude larger than the characteristic van der Waals energy (defined as $E_6=\hbar^{2}/2\mu R_6^{2}$, where $R_6=(2\mu C_6/\hbar^{2})^{1/4}$ is the characteristic distance). 
As a result, all the MQDT functions of the exit channels take their $E\rightarrow\infty$ limit: $C(E)\rightarrow 1$ and $\tan\lambda(E)\rightarrow 0$. 
Additionally, any possible couplings between partial waves will be neglected as they are not known.
They could be taken into account in numerical calculations by diagonalising the interaction matrix at short range before imposing MQDT boundary conditions and using the resulting adiabatic potentials curves.
It can also safely be assumed that the exit channels are not coupled, and the $\mathbf{Y}$ matrix can be chosen in the form
\begin{equation}
\mathbf{Y}=\left(
\begin{array}{lll}
0&y_1&y_2\\
y_1&0&0\\
y_2&0&0
\end{array}
\right),
\end{equation}
where $y_1$ and $y_2$ depend weakly on energy and partial wave quantum numbers. 
The diagonal terms of $\mathbf{Y}$ can be chosen to be zero provided that the reference potentials reproduce the actual phase shifts $\xi_i$ in each of the channels.

The total loss rate from the entrance channel is defined as
\begin{equation}
\mathcal{K}_{0}^\mathrm{re}(E)=\sum_{\ell,m}{\mathcal{K}^\mathrm{re}_{0\ell m}(E)} = \frac{h}{2\mu k} \sum_{\ell,m}{\left(1- |S_{0\ell m,0\ell m}(E)|^2\right)},
\end{equation}
where the entrance channel is labelled by the index $0$ and the exit channels are labelled by indices $1$ and $2$. 
Then, by unitarity of the $\mathbf{S}$ matrix, 
\begin{equation}
1- |S_{0\ell m,0\ell m}(E)|^2=|S_{0\ell m,1\ell m}(E)|^2+|S_{0\ell m,2\ell m}(E)|^2.
\end{equation}
The total ion production rate can be divided into contributions from the two product channels. 
Under all the assumptions made, the off-diagonal elements of the $\mathbf{S}$ matrix can be found using equations~\eqref{SMQDT} and~\eqref{RMQDT}, yielding
\begin{equation}
S_{0\ell m,j\ell m}=-\frac{2 C_0(E,\ell) e^{i(\xi_0+\xi_j)}y_j}{i(y_1^{2}+y_2^{2})+C_0(E,\ell)^{2}\left(i+\tan\lambda_0(E,\ell)\left(y_1^{2}+y_2^{2}\right)\right)}.
\end{equation}

From this one can calculate the branching ratio $\Gamma=\left|S_{01}\right|^{2}/(\left|S_{01}\right|^{2}+\left|S_{02}\right|^{2})$, which reduces to 
\begin{equation}
\Gamma=\frac{y_1^{2}}{y_1^{2}+y_2^{2}}.
\end{equation}

This shows that energy- and angular momentum independent MQDT predicts constant branching ratio. 
However, although the quantum defect parameters $y_i$ have weak energy dependence, especially at kinetic energies much lower than the energy scales characteristic for rovibronic or electronic excitations of the collision complex, one cannot assume that they stay constant over the whole energy range covered in the present experiment which spans over three orders of magnitude. 
Corrections to $y$ for higher partial waves, which can be very different for the two reaction channels, are also expected to influence the branching ratio.

As discussed above, the dominating term in the long range potential is the van der Waals interaction. 
If a pure van der Waals force is considered as the reference potential then the reactive rate constant at energies $E\gtrsim 100E_6$ agrees with the Langevin capture model.\cite{Langevin1905,Jachymski2013} 
This model bases on the assumption that all classical trajectories that fall into the collision center contribute to the reaction rate with equal probability $P_{re}$.
The Langevin capture rate constant $k_{LC}$ is given by
\begin{equation}
k_{LC}(E_{coll}) = P_{re}\frac{6\pi}{\sqrt{2\mu}}\left(\frac{C_6}{4}\right)^{1/3}E_{coll}^{1/6} \label{eq:Langevin},
\end{equation}
and has a power-law dependence in the collision energy. 
It can be shown that $P_{re}=\frac{4t}{(1+t)^{2}}$, where in the present case $t=y_1^{2}+y_2^{2}$.\cite{Jachymski2013} 
Since the total probability cannot be larger than one this sets boundaries on the $y$ parameters. 

One can expect that at very high energies (much larger than the characteristic van der Waals energy, and rather comparable to the potential well depth) the particles will start probing the short range regions of the potential, where deviations from van der Waals are large and higher order terms are important. 
In this case the Langevin model is not sufficient to describe the reaction rates. 
For higher partial waves the interaction can even be totally repulsive due to finite depth of the potential well. 
The behavior of the rate constant in this case changes from power-law to logarithmic.\cite{Dalgarno1950s,Cote2000} 
To account for this in the model, a $6-12$ Lennard-Jones potential was used in the numerical calculations, setting the MQDT boundary conditions at the bottom of the well (for pure van der Waals the starting point does not influence the result as long as it is chosen behind the centrifugal barrier, but in the present case it does matter). 
It should be noted that the additional $\propto r^{-12}$ term is a phenomenological correction neccessary to account for the short range repulsive core, but could also be chosen to have other form.

\section{RESULTS AND DISCUSSION}
\begin{figure}
 \includegraphics[width=\figwidth]{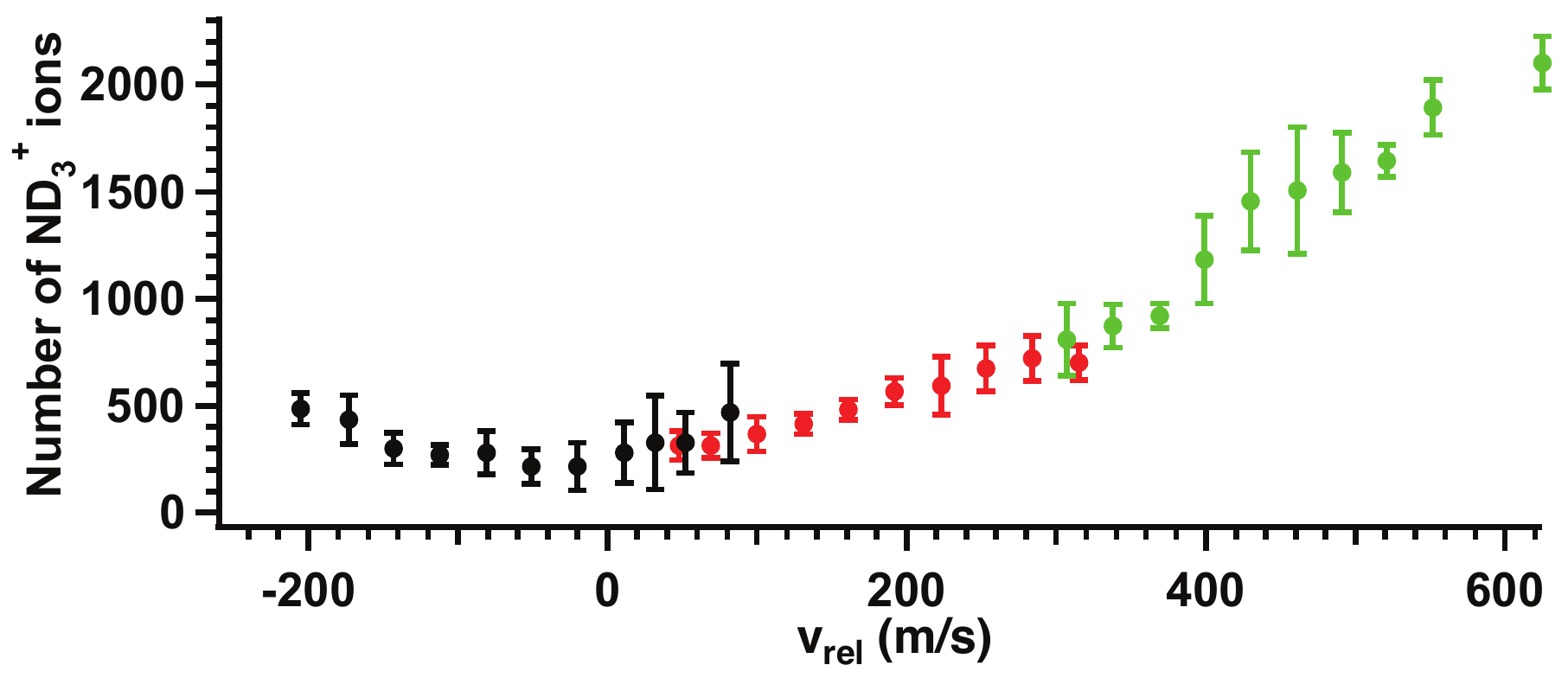}%
\caption{\label{alldata}Relative reaction rates normalised for reactant densities for the Ne*+\nhds reaction as a function of collision energy. Black symbols show data points recorded when operating SE2 with an \ndd/Ar mixture, the red ones with an \ndd/Ne mixture, and for the green symbols neat ammonia was used.}
 \end{figure}
Figure \ref{alldata} shows the density-normalised number of \nddps ions acquired from the $\mathrm{Ne}^*$ + \ndd $\,\rightarrow$ Ne + \nddp\, reaction as a function of relative velocity, $v_{\mathrm{Ne}^*}-v_{\mathrm{ND}_3}$.
The data are sectioned in three parts, where the black symbols show data points recorded SE2 was operated with an \ndd/Ar mixture, the red ones with an \ndd/Ne mixture, and for the green symbols neat ammonia was used.
For each type of ammonia expansion the temperature of SE2 was tuned over the range of 200 K-340 K.
Since the normalisation was done by using the measured densities in the Ne and \ndds beams, according to equation \ref{eq:rate} the data in figure \ref{alldata} effectively represent a relative rate constant.
The ion numbers in figure \ref{alldata} were corrected for the measured beam densities but not scaled any further, demonstrating that the normalisation procedure provides reliable numbers also for different expansion conditions.
Because the rate constant depends on the collision energy, which in turn is given by $E_{coll} = \frac{\mu v_{rel}^2}{2}$, data like those shown in figure \ref{alldata} have to be symmetric about $v_{rel}=0$.
Traces like figure \ref{alldata} are a useful check for the experiment that confirm that within the error margins all the velocity settings are correct.

\begin{figure}[ht]
 \includegraphics[width=\figwidth]{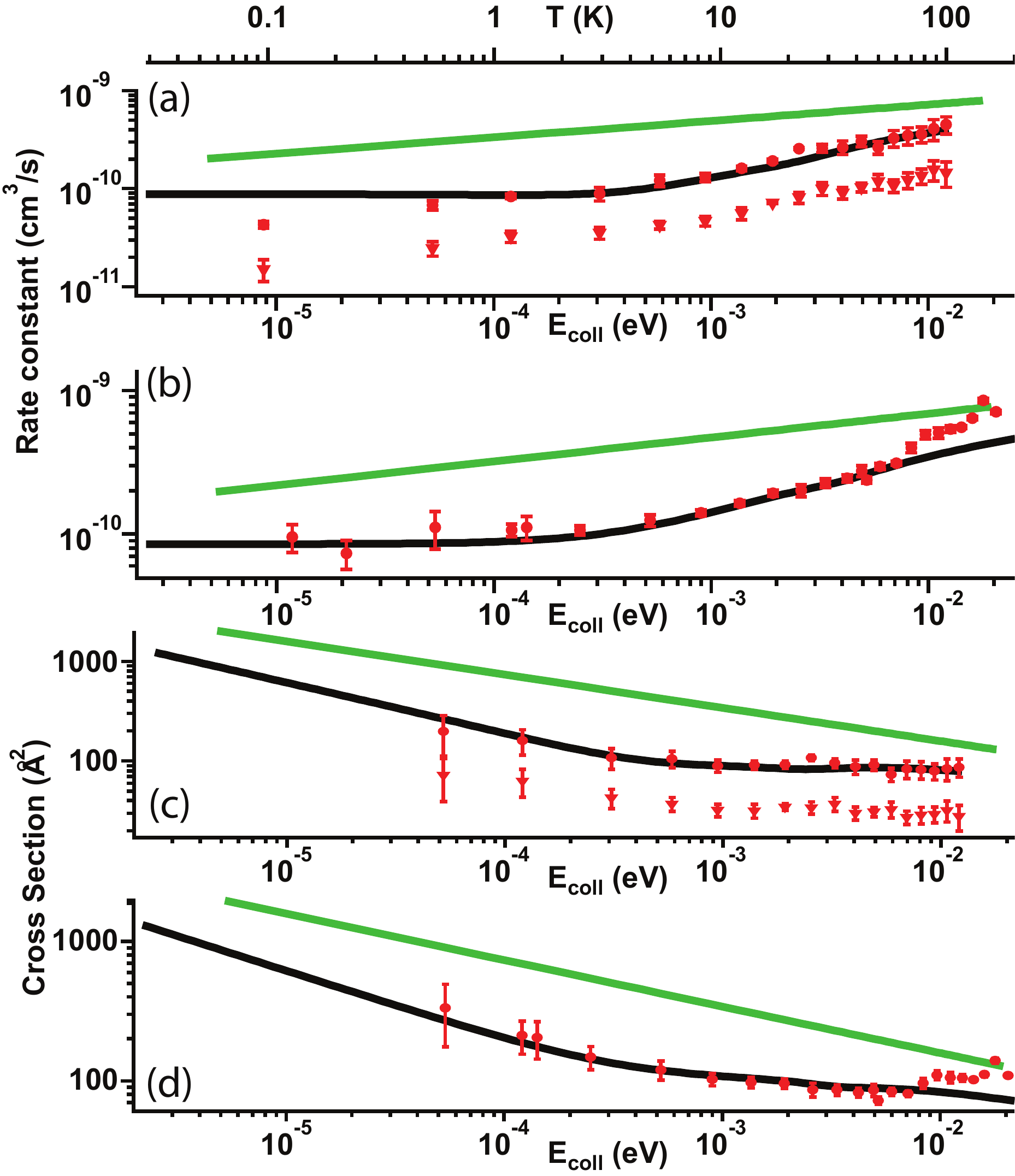}%
\caption{\label{cross-sections}Rate constants for (a) the Ne$^*$+\nhds reaction and (b) the Ne$^*$+\ndds reactions. In both panels the red symbols are the experimental data and the black lines are the results from the MQDT calculations. 
The green lines show the rate predicted on the basis of a Langevin model with unit reaction probability at short range. In panel (a) the circles (triangles) show the cross section for the channel producing \nhdps (\nhtp) products.
Panels (c) and (d) show the cross sections for Ne$^*$+\nhds and Ne$^*$+\ndd, respectively. 
Symbols and colours like in panels (a) and (b).}
 \end{figure}
The calculated and measured rate constants for the $\mathrm{Ne}^*$ + \nhd\, and $\mathrm{Ne}^*$ + \ndd\, chemical reactions are shown in figures \ref{cross-sections}(a) and (b). 
Panel (a) shows the results for \nhd, panel (b) for \ndd.
Panels (c) and (d) show the cross sections for the two reactions, using the same graphical representation as in panels (a) and (b).
In the bottom two panels the data points at the lowest collision energies were dropped because the low relative velocities would have produced error bars that are larger than the actual value.
In each panel the symbols are the experimental data and the black line the result from the MQDT calculations.
In panels (a) and (c) the circles show the rate constant measured for \nhdps production and the triangles that for \nhtps production.
The experimental data for \nhdps and \nddps production were scaled to fit the theoretical data and in this way brought on an absolute scale.
Error bars represent the statistical fluctuations obtained throughout the repetition of the experiment at each collision energy.
For both the \ndds and the \nhds reactions Lennard-Jones potentials with the same van der Waals coefficient and adjustable $C_{12}$ are used for the theoretical modelling.
The $C_{12}$ coefficient connects to the resulting well depth of the potential via $C_{12}=C_6^2/4D$.
For \nhds a well depth of $D=1.1$ K is assumed, and for \ndds it is $D=2.5$ K.
These values are not the true minima of the Ne*-ammonia potentials but are merely numerical fit parameters.
We stress that they are not related to the minima obtained using SAPT.
The green lines in figure \ref{cross-sections} result from a Langevin capture model for a $V(r)\propto r^{-6}$ long range potential with unit short range reaction probability.
The Langevin capture model model predicts a linear relationship between log(k) and log($E_{coll}$).
Clearly, the observed rates at high energies show a behaviour that strongly deviates from the prediction by the capture model.
On the other hand the MQDT calculations reproduce the measured data very well.
They nicely show the transition from the low-energy collision range where the interaction is dominated by the long range potential to the high-energy range where the short range interaction dominates.
The theoretical data in this plot have been convoluted with the experimental resolution of $\pm$15 m/s, and thus no resonances are visible.
The raw calculations do show resonances with positions that are defined by the reference potential, but given that their position cannot be related to the experimental data they are not shown here.

\begin{figure}
 \includegraphics[width=\figwidth]{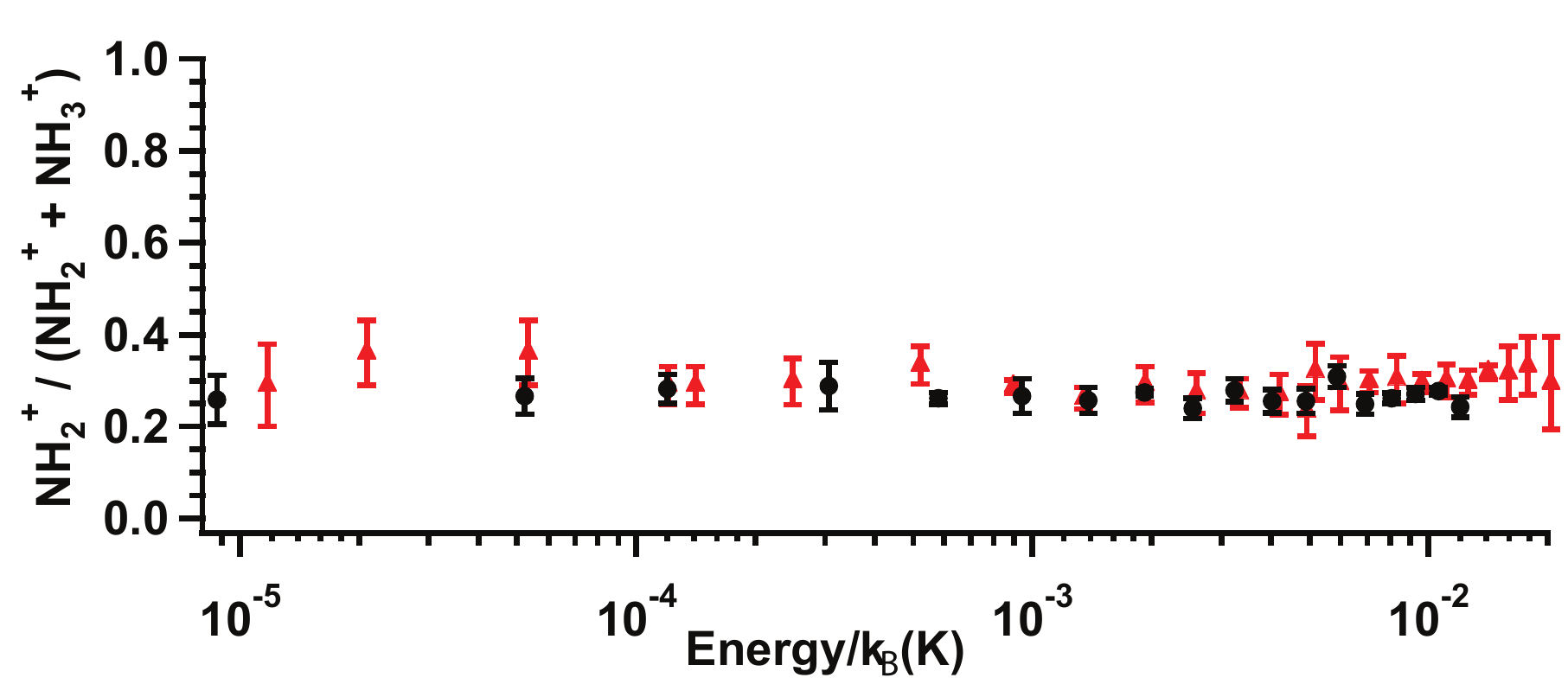}%
\caption{\label{branching}Experimental branching ratios $\Gamma=\frac{[\mathrm{NX}_2^+]}{[\mathrm{NX}_2^+]+[\mathrm{NX}_3^+]}$ (X=H or D) as a function of collision energy. Black circles show the results for the Ne*+\nhd, red triangles for Ne*+\ndd.}
 \end{figure}
 
In addition to the reaction rate constants for equations \ref{eq:nodiss} and \ref{eq:diss}, the experiment provides information on the branching ratio between the two accessible channels.
Figure \ref{branching} shows the results obtained for \nhds (black circles) and \ndds (red triangles), respectively.
While at at higher collision energies the branching ratio for \nhds varied with collision energy,\cite{BenArfa:1999el} the branching ratios $\Gamma$ measured here are both almost constant, $\Gamma\approx0.3$, over the entire range of collision energy even though the measurements span over three orders of magnitude.
This result agrees with the simplest form of the MQDT reaction model, but that model assumes energy-independent quantum defects that at this point can not be justified (this would require a comparison to calculations involving a full PES).
Furthermore, the model would predict the branching ratio to be completely energy independent which is clearly not the case since Ben Arfa et al.\cite{BenArfa:1999el} measured a different value than is found here, with a larger fraction of NH$_2^+$, and a weak energy dependence.
We at this point do not feel that we can give a satisfactory explanation for the observed branching ratio.
Only tentative qualitative arguments that might partly explain the results are presented here.

In view of the arguments given in the introduction one might conclude that the MEP should lead to a dynamic orientation of the ammonia molecules as they approach the Ne*.
There is no full PES available for the Ne*-ammonia complex but in analogy to other, chemically similar complexes, as well as on the basis of our own SAPT calculations one can see that there is a deep well as the Ne* binds to the lone pair of ammonia. 
In the case of the Ne*-water complex, for example, the well along the O-lone pair --Ne* coordinate is more than 400 meV deep while there is barely any minimum along the O-H--Ne* coordinate.\cite{Brunetti:2013}
Similarly, our own calculations predict a minimum of 860 meV in the N-lone pair --Ne* coordinate vs. only 4.5 meV along the N-H--Ne* coordinate.
Accordingly, one might expect the branching ratio to be strongly in favor of formation of \nhdps rather than \nhtp, at the low end of the collision energy spectrum.
This is not observed.
One might conclude that the well along the favourable reaction channel is sufficiently deep in the case of \nhds to completely determine the branching ratio already at $E_{coll}$ as high as 250 K. 
From previous data it is known that all of the \nhdps formed in the ground state remains stable and is detected as \nhdps rather than \nhtps.
The branching for the excited state of \nhdps to dissociate into \nhdps and \nhtps can be assumed to be independent of collision energy because the variation of the collision energy is small compared to the relevant internal energies.
It must thus be concluded that either there is a 30\% probability for formation of the excited state even when the Ne* approaches along the lone-pair axis, or the ammonia molecule indeed does not align at all.
Based on the following arguments we conclude that the naive interpretation of the MEP can clearly not be applied to the current system.

For an alternative explanation of the results, the influence of molecular rotation at different collision energies is considered in more detail .
An approximate reaction time $t_c$ is defined as
\begin{equation}
t_c = \frac{b_c}{v_{rel}}, \label{eq:tc}
\end{equation}
where $b_c$ is a critical impact parameter, i.e. the internuclear separation below which the opacity function, or reaction probability, is unity. 
If $b_c\approx\sqrt{\sigma}$ is assumed, then from the above MQDT calculation $b_c\approx$10 \AA, and 25 \AA\hspace{1 mm} are obtained at the very highest and lowest $E_{coll}$ values studied (corresponding to velocities 600 m/s, and 15 m/s), respectively.
Equation \ref{eq:tc} yields $t_c(E_{coll}= 8 \mu$eV) = 170 ps, and $t_c(E_{coll}$ = 22 meV) = 1.7 ps. 
This reaction time can be compared with the rotational period $t_{rot}$ which is $t_{rot}\approx$3 ps for the \jk{J}{K}=\jk{1}{1} state of \ndd.
It may also be useful to look at the de Broglie wave length of the reactants.
The Ne--\nhd$^*$ system has an associated $\lambda_{dB}$ of 18 \AA, and 0.4 \AAs at a collision energy of 8 $\mu$eV, and 22 meV, respectively. 

Based on this one can define two principal domains: in the first one $t_c\ll t_{rot}$, while in the second one $t_c\gg t_{rot}$.
Since short collision times are associated with high relative velocities the first domain also involves short de Broglie wave lengths while they are large in the second domain.
The present experiments cover the second domain while the first domain has been covered by Ben-Arfa et al.\cite{BenArfa:1999el} 
There, the authors have argued that the almost constant fraction of \nhtps products with increasing collision energy is due to a more statistical nature of the Ne\nhd$^*$ complex at the turning point: the orientation of the ammonia molecule in the complex is random because the Ne* approaches on a much faster time scale than the rotation of the molecules ($t_c\ll t_{rot}$).\cite{BenArfa:1999el} 
In contrast, when $t_c\gg t_{rot}$ the rotation of the ammonia is much faster than the reaction time and the model proposed by Ben-Arfa et al. cannot be applied.
In addition, the long de Broglie wave lengths make the classical picture of the structure of the ammonia molecule questionable.
In this energy range the dynamics must be visualised differently: since the ammonia is rotating very quickly the approaching Ne* atom does not perceive the explicit structure of the molecule.
Much rather does it interact with some averaged, almost spherical object.
Due to the presence of the permanent dipole moment of the ammonia, however, this object can not be assumed completely isotropic.
Nevertheless, the branching ratio again becomes independent of collision energy since the propensity for production of the ion in the ground or excited state only depends on the relative sizes of the cones of acceptance.

This model raises two questions: why are the branching ratios in the two regimes different, and what is the branching ratio and its energy dependence in the regime $t_c\approx t_{rot}$?
Since the present studies do not extend to the lowest collision energies covered by Ben-Arfa et al. it is unfortunately not possible to answer the second question.
Furthermore, the branching ratio is indeed not completely constant in the high-energy range which could either mean that the above model is insufficient to describe the dynamics or the relevant energy range starts at higher collision energies only.
The first question can, however, be discussed qualitatively.
The statistical nature of the process is the same if the ammonia is rotating fast and the direction of the incoming Ne* is constant or if the ammonia is completely fixed in space but different approaches of Ne* are compared.
Based on the arguments given above one would thus expect the branching ratio to be the same at very high and at very low collision energies.
But in different energy regimes the reactants probe different regions of the interaction potential.
At low energies the dynamics depend principally on the long range forces.
The observed interaction potential then contains the isotropic dispersion interaction and the partially anisotropic dipole-induced dipole interaction, which both depend on $r^{-6}$.
At short range different interaction terms become dominant and it can be expected that the anisotropy in the potential also changes its nature.
As a result, the degree of isotropy of the interaction potential is effectively energy dependent.
This would support the conclusion that the branching ratios must be different at low and at high collision energies, because the relative size of the cones of acceptance reflects the different anisotropies of the potential at short range and at long range.

The above model appears to describe the observations qualitatively but leaves some fundamental questions open.
It should also be noted that the time scales defined based on rotational periods are only approximate, in particular because for the rotational period only the lowest state is assumed, and the presence of rotationally excited states with higher rotational speeds is neglected.

Constant branching ratios between different reaction channels have also been observed in other Penning ionization experiments at collision energies above room temperature. 
Vecchiocattivi and his co-workers have found that collisions between $\mathrm{Ne}^*$ and water molecules yield H$_2$O$^+$ products in ground and first electronically excited states, and the fraction of electronically excited H$_2$O$^+$ ions decreases modestly with increasing collision energy.\cite{Balucani:2012wv}
This behaviour is opposite to that seen in the $\mathrm{Ne}^*$ + \nhds reaction in the range of $E_{coll}$ between 40 and 400 meV. \cite{BenArfa:1999el}
Similarly, the fraction of electronically excited N$_2$O$^+$ Penning ions decreases with increasing collision energy, as found in the $\mathrm{Ne}^*$ + N$_2$O reaction.\cite{Biondini:2005bw}
The authors of the above studies did not propose any viable mechanisms that would explain the trends exhibited by the branching ratio as a function of the collision energy. 
More work from the Perugia laboratory on the $\mathrm{Ne}^*$ + CH$_3$X (X = Cl, Br) reaction has shown that the fraction of CH$_3^+$ and CH$_2$X$^+$ products grows with increasing collision energy, whereas the amount of CH$_2$X$^+$ parent ions declines at higher $E_{coll}$ values.  
This was attributed to a softer repulsive wall around the methyl group. 
A recent study conducted in our lab showed almost constant branching ratios for the $\mathrm{Ne}^*$ + CH$_3$F reaction in the same energy range as covered in the present work.\cite{jankunas:arxiv}
Since the rotational period of the heavier CH$_3$F is somewhat lower than for ammonia this may not be a surprise in view of the explanation offered above.
Note that the CH$_3^+$ and CH$_2$X$^+$ products come primarily from collisions between $\mathrm{Ne}^*$ and the methyl end of CH$_3$X molecule. 
The hypothesis was given support by virtually constant branching ratios for the $\mathrm{Ne}^*$ + CH$_4$ Penning ionization. 
The relative fraction of CH$_4^+$, CH$_3^+$, and CH$_2^+$ products did not change in the collision energy range from 35 meV to 300 meV.   

\section{Conclusions}
The present experiments for the first time give access to the reactivity of a polyatomic molecule below 1 K.
The Penning Ionization process investigated here shows a clear transition from a process that is dominated by long-range forces at low collision energies to one where the short range forces define the reaction rate at collision energies above $\approx$100 K.
The entire range can be well described by a MQDT model in which the long range forces are expressed using only van der Waals induction and dispersion forces. 
The short range dynamics are only reproduced satisfactorily if in addition to the pure van der Waals potential a repulsive component of a Lennard Jones potential is included.

Two reaction channels were observed for each of the isotopologues NH$_3$ and ND$_3$. 
No clear isotope effect has been observed.
PI can proceed either through a dissociative channel or through a non-dissociative one.
Based on previous observations these channels are interpreted as different orientations of the ammonia molecules during the reaction.
Within the framework of the minimum energy path one might expect a strong dependence of the branching ratio on collision energy, given that the energy of the Ne*-\nhds complex must be assumed to strongly depend on the  ammonia orientation.
In contrast to this prediction the results show, within the error bars, constant branching ratios over the entire range of collision energies covered here.
A complete and quantitative explanation of the branching ratio will require detailed calculations of the potential energy surface which at this point appear to be out of reach, and the arguments remain qualitative.

\section{Acknowledgments}
Support from the Swiss National Science Foundation (grant number PP0022-119081) and EPFL is acknowledged. 
K.J. was supported by the Foundation for Polish Science International PhD Projects Programme co-financed by the EU European Regional Development Fund. 
M.H. was supported by the project ''Towards Advanced Functional Materials and Novel Devices: Joint UW and WUT International PhD Programme'', operated within the Foundation for Polish Science MPD Programme co-financed by the EU European Regional Development Fund and by the Polish Ministry of Science and Higher Education Grant No. N204 248440. %\bibliography{Allrefs}

\end{document}